# HowTo Authenticate and Encrypt

**by
Hans-Rudolf Thomann[1]**

| | |
|---|---|
| **Author** | Hans-Rudolf Thomann |
| **File name** | HowTo Authenticate and Encrypt.doc |
| **Version** | 1.0 |
| **Date** | 2006-11-02 |
| **Status** | Final |
| **Action** | for consideration |
| **Distribution** | arXive.org |

---

[1] Im Altried 1h, CH-8051 Zürich, hr@thomannconsulting.ch

# CONTENTS

# 0      Abstract

Recently, various side-channel attacks on widely used encryption methods have been discovered. Extensive research is currently undertaken to develop new types of combined encryption and authentication mechanisms. Developers of security systems ask whether to implement methods recommended by international standards or to choose one of the new proposals. We explain the nature of the attacks and how they can be avoided, and recommend a sound, provably secure solution: the CCM standard.

Keywords: authentication, encryption, side-channel attacks

# 1 Introduction

Recently, various side-channel attacks on symmetric encipherment schemes have been discovered. These attacks exploit linear properties characteristic for most symmetric stream ciphers internationally standardized in conjunction with some encodings and padding methods. Authenticated encryption has been concluded to be the soundest way to prevent such attacks once and forever.

Current research thus focuses on the design of mechanisms providing authentication and encryption at once. Though significant progress has been made, there is yet no globally accepted solution, let alone one internationally standardized. Any solution, before it can be globally accepted, must undergo extensive analysis and testing by the global cryptographic community. This procedure took almost ten years for the AES, and it will take several years for new authenticate-and-encipher mechanisms as well. At best such new mechanism can be expected in the next revisions of the international standards.

The many results published make it difficult for practitioners to get a clear picture and to come to sound conclusions. Designers of security systems are facing the question whether to implement those mechanisms recommended by international standards or to trust one of the new proposals. On the one hand, compliance with standards has interoperability advantages and may be a demand for legal reasons. On the other hand, nobody would build a new system on building blocks with known security flaws. But although new proposals promise to resolve the issues, they bear the risk of unknown weaknesses and thus are not viable alternatives before their international approval.

However, CCM, a new standard implementing authenticated encryption, has already been internationally approved. CCM provably resolves the problems and thus can safely be implemented in the next generation security systems.

## 2      Side-channel Attacks

### 2.1    Introduction

Side-channel attacks are attacks using information that is unintentionally disclosed. Attacks of this kind are systematically studied since the mid 90ies. Today numerous attacks are known (see [8] for a review), many of them very dangerous.

Side-channel attacks on symmetric [1, 6, 7, 9, 14, 19, 20] and asymmetric [10, 11, 12, 13, 26] encryption algorithms have obtained attention since about ten years. A wealth of such attacks is described in the literature (a few examples given above). An overview over padding attacks is given by [19]. We will give a very general description of encipherment and decipherment mechanisms in section 2.2. and sketch some attacks in section 2.3.

The safest way to prevent these attacks is to combine encryption with authentication. Research is therefore focusing on the development of new combined authenticated encryption security mechanisms [15, 16, 17, 18, 20, 21]. Overviews are given by [21] and slightly more detailed [20]. We will explain the effect of authentication in section 2.4.

The new insights have entered into recent revisions of international standards [2, 3, 22], though the innovation process is still ongoing and final, globally approved results are not yet available. We will describe the CCM authenticated encryption mechanism [25] and the CCMP protocol standard [22, section 8.3.3] in section 3, point out some finesses that must be considered and express our recommendation for next generation systems to use CCM, whose security has been proven in [23] and confirmed in [20, 21].

### 2.2    Encipherment and Decipherment

Encipherment and decipherment conforms to a very general scheme. Encipherment mechanism take a plaintext P as input and produce a ciphertext C as output. Most mechanisms do this in two steps. First, they bring P in a certain format, using a formatting function F and producing the intermediary text T. On T they apply the cryptographic transform E yielding C:

Encryption

$$F : P \mapsto T$$
$$E : T \mapsto C$$

For decryption, the decryption transform D reproduces T from C, and format validation V is performed, yielding "INVALID" if T is (for whatever reasons) not well-formatted, whereas reproducing P if everything is ok.

Decryption

$$D : C \mapsto T$$
$$V : T \mapsto \begin{cases} INVALID & (if\ T\ is\ not\ well-formatted) \\ P & (otherwise) \end{cases}$$

The reason for this complication is, that, while encipherment algorithms are bijective functions, formatting rules are not generally surjective. Prominent examples are:

- padding functions, extending an arbitrary bit string to a multiple of octets or blocks

- encoding functions, such as ASCII, Base64, BER-TLV
- checksums, such as checkdigits and CRCs, appending redundancy to a text

## 2.3 Attacks on Decipherment Mechanisms

The most commonly used encipherment mechanisms are the modes of operation standardized in [3], namely CFB and OFB mode, as well as the CBC mode standardized in [24]. These mechanisms all share the property, that for decipherment the ciphertext C is XORed with the keystream Z to obtain the intermediary text T:

$$C \otimes Z \mapsto T$$

This particularity allows for two kinds of attacks.

### 2.3.1 Manipulation of Message Contents

Though these mechanisms perfectly hide the plaintext from an eavesdropper, they do not protect it against manipulation[2]. An attacker intercepting the ciphertext C can modify some bits of C und thus effect a modification of the text T.

Let's now assume that the attacker has sufficient knowledge of T, e.g. he knows that some bit of T expresses a "no" response on an authorization request, then he can turn this into a "yes", just by XORing the ciphertext with "yes" XOR "no".

### 2.3.2 Disclosure of Message Contents

If the attacker knows nothing about T, then from C alone he cannot gain information about T. But, using the technique described in the previous section, he can manipulate T and learn from the behavior of the receiver if the manipulations result in valid or invalid texts. The literature calls this validation attacks, wherein the receiver acts as a validation oracle. If the formatting rule F has certain properties, then a validation attack may yield significant information about T and thus P.

Indeed when CBC mode with padding rule 3 from [5], or rule 1 in conjunction with a length parameter, is used for encipherment, then the attacker can completely recover the plaintext P with only few trials, as shown e.g. in [6]. If padding rule 2 is used (sometimes called OZ padding, as a single 1-bit and 0-bits up to the next block boundary are appended), then [19] proves that no attack is possible, unless the attacker has some prior knowledge of the plaintext, which however is the case in many practical scenarios.

Thus side-channel information about the receiver's format validation result in conjunction with contents manipulation allows a partial or total breach of confidentiality. This situation is very unsatisfactory, as encodings, formatting rules and paddings are not generally considered security-relevant.

## 2.4 Adding Authentication

The above attacks did succeed because there was no message integrity. So let's protect it by adding authentication. However, as we are going to see, lack of integrity protection is not the primary and only deficiency. The order of operations is as well critical.

### 2.4.1 Krawczyk's Attack Scenario

[1] describes an attack on a specific encryption scheme which is perfectly secure against chosen plaintext attacks because it relies on a one-time pad.

---

[2] They do not claim to provide anything else than confidentiality.

The encryption mechanism $E^*$ is defined in [1] as follows: Given an n-bit plaintext x, $E^*$ first applies the encoding F below followed by the encryption function E, i.e.

$$E^*(x) = E\big(F(x)\big).$$

The encoding F is defined by bitwise application of the transform[3]

$$0 \mapsto 00$$
$$1 \mapsto 01 \,\big|\, 10$$

where | denotes a randomized choice. The bit pair $11$ is not a valid encoding result. On decoding it yields "invalid".

[1] describes the following attack when only encryption $E^*$ is used: The attacker A flips two ciphertext bits and sends the modified ciphertext c' to the destination[4]. If A obtains information on whether the decoding of c' yielded a valid result or not, then A can determine the value of the bit whose encoding was modified.

[1] then considers the case where the message is authenticated by appending a MAC M and afterwards encrypting by the function E*. The receiver will first decrypt the message, obtaining sometimes INVALID, sometimes a valid plaintext. In the latter case the MAC is verified. As one easily sees, nothing changes by adding authentication: The attack still works as before.

Thus the source of trouble seems to be the INVALID outcome of the format validation step during decryption. But this is only half of the truth, as we will learn from the next scenario.

### 2.4.2 Improved Attack Scenario

Let's improve the above attack scenario by a slightly different encoding:

$$0 \mapsto 00 \big| 01 \big| 10$$
$$1 \mapsto 11$$

The attacker again switches two bits. But now, the encodings do not any longer become invalid. Thus, this encoding combined with one-time pad encipherment, as in the previous section, is perfectly secure. But when adding authentication, it becomes insecure, as in one half of the cases the plaintext will be the same, in the other half different!

You can do even better: First switch the first bit and submit the message to the receiver. If he accepts, then you know that you got a zero bit. If he rejects, then take the original ciphertext and switch the second bit. If it is rejected as well, then the bit is one else zero.

Here the addition of authentication makes things worse! Format validation does not raise any rejections, but authentication does. What is common to the previous scenario is, that some manipulations do change the plaintext, others don't. And authentication perfectly detects these conditions!

---

[3] [1] claims such encodings to be in use by common encryption schemes. However, [3] only recommends padding, [2] redundancy addition (see e.g. appendix D).

[4] Here [1] tacitly assumes that E have a linear output function, which is the case for one time pads (OTP) and various other stream ciphers, but does not hold in generality. See [3] for a general outline.

 

### 2.4.3  Secure Scenario

The above scenarios are somewhat unusual, as the MAC is computed from plaintext P, the result is formatted with formatting rule F and then enciphered with transform E.

More common is the following order: Take the plaintext P, create the formatted text T, calculate the MAC M from T and append it to T, finally create the ciphertext C from T+M.

As all standardized encipherment functions are bijective, any manipulation of C will cause a change of T+M on the receiver side. Thus MAC verification will always fail. Any manipulated message will be rejected. The attacker does not experience any differences and does not learn anything. The scheme is perfectly secure[5]!

In the terminology of [1] this is the AtE scheme. [1] promotes another scheme, EtA, and proves its security.

### 2.4.4  Conclusion

Let us be slightly more precise and indicate the formatting step by the letter F, authentication by A and encryption by E, and summarize our findings in a table:

| Scenario | Operations | Secure |
|----------|------------|--------|
| section 2.4.1 | A F E | no |
| section 2.4.2 | A F E | no |
| section 2.4.3 | F A E | yes |
| EtA of  [1] | F E A | yes |

As formatting of the ciphertext does not make much sense, F can only be at the first or the second position. The table thus contains all reasonable orders of operations. The secure versions are those starting with the formatting F.

Evidently, for security

- F must come before A and E.
- The order of A and E does not matter[6].

The cryptographic transforms must be applied to the *formatted* text T, not to the plaintext P.

### 2.5  Authenticated Encryption

As we have seen, remedy is found by applying authentication and encryption to the *formatted* text T. The idea is that authentication make manipulations of the attacker detectable, the receiver thus return "INVALID" whenever the attacker manipulates the ciphertext, for all manipulations without any differentiation, thus the attacker not learn anything at all. Let's make this more precise.

Authenticated Encryption

$$F : P \mapsto T$$

$$A : T \mapsto M$$

$$E : T + M \mapsto C$$

---

[5] As proven in [1] and [19].
[6] For other reasons, authentication usually comes before encryption.

Firstly, the plaintext P is formatted with the formatting rule F, yielding the formatted text T. From T the authentication mechanism A computes the MAC M. The MAC is appended to T (the + sign denotes concatenation) and enciphered with the encipherment mechanism E, yielding the ciphertext C.

The receiver of the ciphertext C first decrypts it, obtains T and M, verifies the authenticity of T by re-calculating the MAC and comparing it with M, validates the format and, if it is valid, recovers the plaintext P. In case of negative authentication outcome he returns "MAC failure", in case of positive authentication the validation result, i.e. either INVALID or P.

<u>Authenticated Decryption</u>

$$D : C \mapsto T + M$$

$$A : T \mapsto M\,'$$

$$V : T \mapsto \begin{cases} INVALID & (if\ T\ is\ not\ well-formatted) \\ P & (otherwise) \end{cases}$$

$$Output \begin{cases} MAC\ failure & (if\ M \neq M\,') \\ V(T) & (if\ M = M\,') \end{cases}$$

Any change of C causes a change of T and/or M. Thus M' will mismatch M with high probability. Therefore the attacker does not obtain any significant amount of information about the plaintext.

The following two precautions are absolutely essential for security:

1. The MAC must be calculated from the *formattet text*, T, but *never* from the unformatted plaintext P. Would one calculate the MAC from the plaintext P, then some changes of C could cause a change of P, some not. The latter would pass the MAC verification step, the former not, and the attacker would again learn something about the structure of P.
2. The receiver must output the validation result (i.e. either INVALID or P) *if and only if* the authentication did succeed, i.e. $M = M\,'$. Otherwise he acts again as a validation oracle.
3. The receiver must avoid any difference in timing between MAC failure and INVALID cases. Otherwise an attacker may infer the attack information from timing. It is advisable to perform both, MAC verification and format validation, in all cases.

Current standards such as [3] and [5] provide independent mechanisms, A and E, for encipherment and authentication. In the future we will see combined authenticated encryption mechanisms, A-E, such as [15]. But the above precautions will remain indispensable. The sender will always apply A-E to T, the receiver apply D-A to T, not to P. And he will always disclose the validation result only in case of positive authentication.

## 2.6   Hints

How then should one treat these error conditions, implementers may ask. Well, as today's communications channels are virtually error-free, any MAC failure is strong evidence for either an attempted attack or a severe, probably permanent malfunction of a system component. An ill-formatted text T with valid MAC indicates a malfunction (e.g. a program bug). In both cases the following steps are advisable:

1. Return an error response to the sender (caution: respect the precautions listed in the previous section).
2. Record the error event in the error log.
3. Either cancel the communications session, or allow for very few (say 3) retries.
4. If re-initializing the session, make sure that new keys with sufficient unpredictability (preferably true randomness, not just pseudo-randomness) are established.
5. Implement an auditing process analyzing the error log, investigating the source of the error and taking appropriate action. This is essential both to repair malfunctioning components and for security. The details of such a process depend on the system. In a Web server it may be appropriate to regularly view error log statistics, having a look at individual log entries only if the statistics are above normal. In a peer-to-peer communication system, every event may be worth investigating.

One more word of caution for those wishing to design their own scheme: Do not replace the function A just by a key-less hash function. Function A must include a cryptographic key, otherwise some attacks are again possible.

## 3    CCM

In the year 2004 the NIST has published an authenticated encipherment mechanism named CCM [25]. The name stands for CTR-mode encipherment and CBC-MAC authentication. This is exactly what CCM does. CTR mode is specified in [24] and CBC-MAC in [5]. Thus CCM uses currently standardized mechanisms as building blocks.

CCM moreover strictly specifies the procedures for authenticated encryption and decryption. This specification fits into the scheme given in section 2.3 above, using CBC-MAC for A and CTR mode encipherment for E, even more restrictive by demanding the use of "INVALID" for both, negative authentication and format errors.

CCM does not restrict the formatting rule F beyond the evident requirement that F must not loose information. Formatting of different plaintexts, P and P', must yield different formattet texts, $T = F(P)$ and $T' = F(P')$:

$$P \neq P' \Rightarrow T \neq T'$$

CMM specifies that the same key K be used for both mechanisms, A and E.

The following details are essential for security:

1. The first block of text T contains a nonce value which must be unique for all messages over the whole lifetime of the key K, and distinct from any counter value.
2. All counter values must be distinct over the whole lifetime of the key K.
3. The MAC is enciphered with the keystream value, $S_0$, derived from the *first* counter value, $Ctr_0$.

CCM has been proven secure in [23].

CCM has recently been adopted by the CCMP specification [22]. CCMP is the CCM protocol, replacing the WEP protocol in wireless local area networks (WLAN). CCMP requires the nonces to be message sequence numbers. The receiver must check these sequence numbers to detect replay attacks. This is another feature that any secure implementation should adopt.

## 4    Conclusion

CCM is well-designed and its security has been well analyzed and proven. It is easily imple-mented and can be combined with various key management schemes. For the next genera-tion systems CCM is the method we recommend. New methods should be adopted once they have received global approval and are incorporated into international standards.